\documentclass[preprint,aps,nofootinbib,showpacs,amsfonts,epsf]{revtex4}

\input{epsf.tex}

\newcommand{\E}{{\cal{E}}}
\newcommand{\A}{{\cal{A}}}
\newcommand{\B}{{\cal{B}}}
\newcommand{\C}{{\cal{C}}}
\newcommand{\II}{{\cal{I}}}
\newcommand{\s}{\sigma}
\newcommand{\I}{{\rm i}}
\renewcommand{\d}{{\rm d}}
\renewcommand{\a}{\alpha}

\newcommand{\be}{\begin{equation}}
\newcommand{\ee}{\end{equation}}
\newcommand{\bea}{\begin{eqnarray}}
\newcommand{\eea}{\end{eqnarray}}
\newcommand{\ba}{\begin{array}}
\newcommand{\ea}{\end{array}}
\def\J#1#2#3#4{{#1} {\bf #2}, #3 (#4)}
\def\PRD{Phys. Rev. D}

\def\PRL{Phys. Rev. Lett.}

\def\APL{Ann. Phys. (Leipzig)}

\def\JMP{J. Math. Phys.}

\def\JHEP{J. High Energy Phys.}

\def\CQG{Class. Quantum Grav.}

\def\ZP{Z. Phys.}

\def\PLA{Phys. Lett. A}
\def\PLB{Phys. Lett. B}
\def\NPB{Nucl. Phys. B}

\def\NC{Nuovo Chim.}

\def\ib{{\it ibid.}}

\begin{document}
\draft
\title{Analogs of the double--Reissner--Nordstr\"om solution\\ in magnetostatics and dilaton gravity:\\ mathematical description and some physical properties}

\author{V.~S.~Manko,$^1$ E.~Ruiz$\,^2$ and J.~S\'anchez--Mondrag\'on$\,^1$}
\address{$^1$Departamento de F\'\i sica, Centro de Investigaci\'on y de
Estudios Avanzados del IPN, A.P. 14-740, 07000 M\'exico D.F.,
Mexico\\$^2$Instituto Universitario de F\'{i}sica
Fundamental y Matem\'aticas, Universidad de Salamanca, 37008 Salamanca, Spain}

\begin{abstract}
In this paper we consider a magnetic analog of the double--Reissner--Nordstr\"om solution and construct the corresponding magnetic potential $A_\varphi$ in the explicit form. The behavior of the resulting solution under the Harrison transformation then naturally singles out {\it the asymmetric black diholes} -- configurations composed of two non--extreme black holes possessing unequal masses, and charges equal in magnitude but opposite in sign -- as its most general subclass for which equilibrium of the black--hole constituents can be achieved with the aid of the external magnetic (or electric) field. We also generalize the double--Reissner--Nordstr\"om solution to the dilaton gravity with arbitrary dilaton coupling, yielding as the result the 4--dimensional {\it double--Gibbons--Maeda} spacetime. The study of some physical properties of the solutions obtained leads, in particular, to very simple formulas for the areas of the horizons and surface gravities. \end{abstract}


\maketitle


\section{Introduction}

As was shown in \cite{GMa,DGe,Emp,LTe,ETe}, static axisymmetric solutions of the Einstein--Maxwell--dilaton theory can be generated from the known stationary solutions of the Einstein equations or directly from the static solutions of Einstein--Maxwell theory. The solution--generating procedure permitting one to do this was described in detail in \cite{LTe,ETe}, and an interesting outcome of its use are various dihole spacetimes describing a dilatonic pair of black holes endowed with electric or magnetic dipole moments (the black--hole constituents carry charges which are equal in magnitude but opposite in sign\footnote{In this sense the solution obtained by Liang and Teo \cite{LTe} is not strictly speaking of the dihole type since in general the charges of the constituents are not equal in absolute values.}). Thus, the Davidson--Gedalin solution \cite{DGe} representing a dilatonic generalization of Bonnor's magnetic dipole solution \cite{Bon} can be interpreted, according to a thorough analysis carried out by Emparan \cite{Emp}, as a pair of magnetically charged extreme Reissner--Nordstr\"om black holes (a dilatonic {\it extreme} dihole). Moreover, Emparan demonstrated that the extremal constituents in Bonnor's and Davidson--Gedalin diholes can be balanced by placing the diholes into the external magnetic field. The non--extremal electric diholes were constructed and studied by Emparan and Teo \cite{ETe}. It has been recently shown \cite{CCM} that the Emparan--Teo solutions can be rewritten in a parametrization involving Komar masses and charges of the constituents as arbitrary parameters; the paper \cite{CCM} also clarifies the construction of magnetic analogs of those solutions.

All the extreme and non--extreme diholes considered in \cite{DGe,Emp,ETe}, in addition to being static and axisymmetric, have one more symmetry -- they are equatorially symmetric, i.e., the corresponding metric functions are invariant under the transformation $z\to-z$ \cite{EMR}, and this is a reflection of the fact that these solutions are composed of identical (up to the sign of the charge) constituents. It would be therefore of interest to go beyond the known families of {\it symmetric} dihole solutions and consider a more general case of {\it asymmetric} black diholes -- the ones consisting of a pair of black--hole constituents with unequal masses but equal and opposite charges. In the present paper we will demonstrate a remarkable general property of such solutions possessing zero total magnetic/electric charge to be able to describe equilibrium configurations of two non--extreme Reissner--Nordstr\"om black holes \cite{Rei,Nor} by placing these into the external magnetic/electric field. This task will be accomplished by us within the framework of the magnetostatic analog of the double--Reissner--Nordstr\"om solution \cite{BMA,Man} by constructing the corresponding magnetic potential $A_\varphi$ whose behavior on the symmetry axis provides one with the restrictions on the charges of black holes defining equilibrium configurations in the presence of the external field. Mention that in the case of extremal black holes with non--zero total charge, the inability of the additional magnetic field of the Bonnor--Melvin type \cite{Bon2,Mel} to regularize the symmetry axis was proved by Liang and Teo \cite{LTe}, and later on \cite{ETe} Emparan and Teo conjectured that the same must be as well true for the non--extreme black holes with unequal charges. In our paper the validity of the latter conjecture will find its justification thanks to the use of the explicit formula for the magnetic potential $A_\varphi$ obtained. Moreover, we shall derive the general condition at which the constituents of asymmetric black diholes will be in equilibrium in the external magnetic field.

The second major objective pursued by the present paper is the construction of the 4--dimensional double--Gibbons--Maeda solution for two arbitrary static dilatonic black holes endowed with electric or magnetic charges. It is very advantageous to have such solution on hand for being able to study analytically the general properties of interacting black holes in the presence of a dilaton field, and the desired solution will be generated by us with the aid of a procedure described by Emparan and Teo in \cite{ETe}. We remark that recently several formal and hardly manageable mathematical relations for some physical properties of the interacting Reissner--Nordstr\"om black holes have been presented in \cite{CMRS}, so in our paper we shall pay special attention to elaborating the thermodynamic characteristics of both the double--Reissner--Nordstr\"om and double--Gibbons--Maeda solutions in a concise explicit form.

The paper is organized as follows. In Sec.~II we review the double--Reissner--Nordstr\"om solution and construct its magnetic analog. The main original results of this section are the remarkably simple formulas for the horizons' areas of interacting Reissner--Nordstr\"om black holes and the analytical expression of the magnetic potential $A_\varphi$ for two magnetically charged static black holes. In Sec.~III we consider a family of asymmetric diholes arising as a special subclass of the double--Reissner--Nordstr\"om solution for which the equilibrium of black--hole constituents can be achieved by means of the Harrison transformation \cite{Har}. Here we find a simple formula for the magnetic parameter of the exterior field which determines equilibrium of two static black holes endowed with unequal masses and equal but opposite charges. Sec.~IV is devoted to the derivation of the dilatonic generalization of the double--Reissner--Nordstr\"om solution which we interpret as a non--linear superposition of two 4--dimensional Gibbons--Maeda solutions \cite{GMa}; in this section we also briefly comment on the physical properties of the double--Gibbons--Maeda spacetime. Concluding remarks are given in Sec.~V.

\section{The double--Reissner--Nordstr\"om solution and its analog in magnetostatics}

Exact electrostatic 5--parameter solution describing the non--linear superposition of two arbitrary Reissner--Nordstr\"om black holes has been worked out in a closed analytical form in the papers \cite{BMA} and \cite{Man} using different but mathematically equivalent parametrizations. Since we are interested in the binary black--hole systems, in what follows we shall write out the double--Reissner--Nordstr\"om solution in the Varzugin--Chistyakov parametrization \cite{VCh} employed in \cite{Man} which involves as arbitrary parameters the relative distance and the individual Komar masses and charges \cite{Kom} of the constituents; we shall also slightly change the notations of the paper \cite{Man} to make them more congruent with the recent physical representation of the Emparan--Teo symmetric dihole solution \cite{CCM}. The Ernst potentials $\E$ and $\Phi$ \cite{Ern} defining the double--Reissner--Nordstr\"om solution have the form \cite{Man} \bea
\E&=&\frac{\A-\B}{\A+\B}, \quad
\Phi=-A_t=\frac{\C}{\A+\B}, \nonumber \\
\A&=&\Sigma\,\s[\nu(R_++R_-)(r_++r_-)+4\kappa(R_+R_-+r_+r_-)] \nonumber\\ &-&(\mu^2\nu-2\kappa^2)(R_+-R_-)(r_+-r_-), \nonumber\\ \B&=&2\Sigma\,\s[(\nu m+2\kappa M)(R_++R_-)+(\nu M+2\kappa m)(r_++r_-)] \nonumber\\ &+&2\s[\nu\mu(Q-\mu)-2\kappa(RM-\mu q-\mu^2)](R_+-R_-) \nonumber\\ &+&2\Sigma[\nu\mu(q+\mu)+2\kappa(Rm+\mu Q-\mu^2)](r_+-r_-), \nonumber\\  \C&=&2\Sigma\,\s\{[\nu(q+\mu)+2\kappa(Q-\mu)](R_++R_-)+[\nu(Q-\mu) +2\kappa(q+\mu)](r_++r_-)\} \nonumber\\ &+&2\s[\mu\nu M+2\kappa(\mu m-RQ+\mu R)](R_+-R_-) \nonumber\\ &+&2\Sigma[\mu\nu m+2\kappa(\mu M+Rq+\mu R)](r_+-r_-), \label{EF} \eea where $A_t$ is the electric component of the electromagnetic 4--potential $A_i=(0,0,0,A_t)$, the functions $R_\pm$ and $r_\pm$ are defined as \bea R_\pm&=&\sqrt{\rho^2+(z+{\textstyle\frac{1}{2}}R\pm\Sigma)^2}, \quad r_\pm=\sqrt{\rho^2+(z-{\textstyle\frac{1}{2}}R\pm\s)^2}, \nonumber \\ \Sigma&=&\sqrt{M^2-Q^2+2\mu\,Q}, \quad \s=\sqrt{m^2-q^2-2\mu\,q}, \quad \mu:=\frac{mQ-Mq}{R+M+m}, \label{Rr} \eea and $R$ is the distance between the centers of black holes; the constant objects $\nu$ and $\kappa$ are given by the formulas \bea \nu&=&R^2-\Sigma^2-\s^2+2\mu^2\equiv R^2-M^2-m^2+(Q-\mu)^2+(q+\mu)^2, \nonumber\\ \kappa&=&Mm-(Q-\mu)(q+\mu). \label{nk} \eea Note that the upper Reissner--Nordstr\"om constituent is described by the functions $r_\pm$ and it possesses the mass $m$, charge $q$ and the irreducible mass $\s$ (see Fig.~1). The location of the lower constituent is determined by the functions $R_\pm$, and its mass, charge and irreducible mass are $M$, $Q$ and $\Sigma$, respectively. The total mass $M_T$ and total charge $Q_T$ of the system are simple sums of the individual masses and charges, respectively: \be M_T=M+m, \quad Q_T=Q+q. \label{MQ_tot} \ee

The metric functions $f$ and $\gamma$ of the double--Reissner--Nordstr\"om solution which enter as coefficients into Weyl's line element \be
\d s^2=f^{-1}[e^{2\gamma}(\d\rho^2+\d z^2)+\rho^2\d\varphi^2]-f\d
t^2, \label{Weyl} \ee are expressible in terms of $\A$, $\B$ and $\C$ as follows:
\bea
f&=&\frac{\A^2-\B^2+\C^2}{(\A+\B)^2}, \quad
e^{2\gamma}=\frac{\A^2-\B^2+\C^2}{K_0^2 R_+R_-r_+r_-}, \nonumber\\ K_0&=&4\Sigma\,\s[R^2-(M-m)^2+(Q-q-2\mu)^2].  \label{fg} \eea

Obviously, formulas (\ref{EF})--(\ref{fg}) describe a pair of interacting non--extreme black--hole constituents when $\Sigma^2>0$ and $\sigma^2>0$.

\subsection{Some properties of the double--Reissner--Nordstr\"om solution}

Since in the paper \cite{Man} the thermodynamical characteristics of the double--Reissner--Nordstr\"om solution were not considered, below we briefly comment on them within the framework of the well--known Smarr's mass formula \cite{Sma} valid for each black--hole constituent: \bea M&=&\Sigma+\Phi_\Sigma Q=\frac{1}{4\pi}\kappa_\Sigma A_\Sigma+\Phi_\Sigma Q, \nonumber\\ m&=&\sigma+\Phi_\sigma q=\frac{1}{4\pi}\kappa_\sigma A_\sigma+\Phi_\sigma q, \label{smar_gen} \eea where the quantities $\Phi_\Sigma$, $A_\Sigma$ and $\kappa_\Sigma$ describing the lower black hole are, respectively, the constant value of the potential $A_t=-\Phi$ on the horizon, the area of the horizon and the surface gravity; the analogous characteristics of the upper black hole are $\Phi_\sigma$, $A_\sigma$ and $\kappa_\sigma$. Using (\ref{EF}), one easily finds that \be \Phi_\Sigma=\frac{Q-2\mu}{M+\Sigma}, \quad \Phi_\sigma=\frac{q+2\mu}{m+\sigma}. \label{fi_gen} \ee At the same time, the calculation of $A_\Sigma$ and $A_\sigma$ which can be carried out for instance with the aid of the formulas \cite{ETe} \bea A_\Sigma&=&4\pi\Sigma(\rho f^{-1} e^\gamma)|_{\rho=0,-\Sigma\le z+\frac{1}{2}R\le\Sigma}, \nonumber\\ A_\sigma&=&4\pi\sigma(\rho f^{-1} e^\gamma)|_{\rho=0,-\sigma\le z-\frac{1}{2}R\le\sigma}, \eea
is by far more difficult since the initial expressions resulting from the straightforward calculations are extremely complicated and unwieldy. Fortunately, after a laborious and prolonged work we have been able to eventually arrive at the following elegant formulas for $A_\Sigma$ and $A_\sigma$: \bea A_\Sigma&=&\frac{4\pi[(R+M+m)(M+\Sigma)-Q(Q+q)]^2}{(R+\Sigma)^2-\sigma^2}, \nonumber\\ A_\sigma&=&\frac{4\pi[(R+M+m)(m+\sigma)-q(Q+q)]^2} {(R+\sigma)^2-\Sigma^2}, \label{ai_gen} \eea and we remark that in obtaining (\ref{ai_gen}) a concise expression for the horizon area of the Emparan--Teo non--extreme dihole \cite{ETe} was a guiding line for us. Lastly, the form of $\kappa_\Sigma$ and $\kappa_\sigma$ can be read off from (\ref{smar_gen}) and (\ref{ai_gen}): \bea \kappa_\Sigma&=&\frac{\Sigma[(R+\Sigma)^2-\sigma^2]}{[(R+M+m)(M+\Sigma)-Q(Q+q)]^2}, \nonumber\\ \kappa_\sigma&=&\frac{\sigma[(R+\sigma)^2-\Sigma^2]}{[(R+M+m)(m+\sigma)-q(Q+q)]^2}. \label{sg} \eea

Formulas (\ref{ai_gen}) and (\ref{sg}) clearly illustrate that the affirmation made in a recent paper \cite{CMRS} about the absence of simple expressions for the surface gravities and horizons' areas in the general case of the double--Reissner--Nordstr\"om solution is not correct.

\subsection{The case of magnetically charged black holes}

Because of the one--to--one correspondence existing between the electrostatic and magnetostatic Einstein--Maxwell fields \cite{Bon2}, the Ernst potentials of the magnetic analog of the double--Reissner--Nordstr\"om solution have the form \be \E=\frac{\A-\B}{\A+\B}, \quad \Phi=\frac{\I\C}{\A+\B}, \label{EF_mag} \ee where the functions $\A$, $\B$ and $\C$ are the same as in formulas (\ref{EF}), and the corresponding metric functions $f$ and $\gamma$ are given by Eqs.~(\ref{fg}). At the same time, finding the remaining magnetic potential $A_\varphi$ [the $\varphi$--component of the 4--potential $A_i=(0,0,A_\varphi,0)$] represents a difficult technical problem because, unlike in the electrostatic case, its form cannot be read off directly from the expression of $\Phi$. Formally, $A_\varphi$ can be obtained by solving the following system of the first--order differential equations: \be \frac{\partial A_\varphi}{\partial\rho}=\I\rho
f^{-1}\frac{\partial\Phi}{\partial z}, \quad \frac{\partial A_\varphi}{\partial z}=-\I\rho f^{-1}\frac{\partial\Phi}{\partial\rho}. \label{AF} \ee However, in practice, the integration of (\ref{AF}) does not look feasible not only in our 5--parameter case but even in the by far simpler case of identical black holes carrying opposite charges \cite{ETe}. The way out of this problematic situation consists of circumventing the resolution of the system (\ref{AF}) by finding $A_\varphi$ as the real part of Kinnersley's potential $\Phi_2$ \cite{Kin} whose construction within Sibgatullin's integral method \cite{Sib} is analogous to the construction of the Ernst potentials $\E$ and $\Phi$. In the paper \cite{RMM} an explicit expression for $\Phi_2$ was obtained in the general case of the analytically extended multi--soliton solution (see formulas (3.13) and (3.14) of \cite{RMM}). Then, setting $N=2$ in the latter formulas and taking into account that in the magnetostatic case the potential $\Phi_2$ is equal to $A_\varphi$ exactly, one is able, after expanding the determinants and performing the reparametrization within the lines of the paper \cite{Man}, to arrive at the desired expression for the magnetic potential. The calculation of $A_\varphi$, combined with the subsequent tedious search for considerable simplifications, has finally led us to the following result:
\bea A_\varphi&=&Q+q+\frac{\II-z\C}{\A+\B}, \nonumber\\ \II&=&4\Sigma\sigma\{[mR(Q-\mu)-\mu(\kappa+m^2)+\mu(Q-\mu)^2]r_+r_- \nonumber\\ &-&[MR(q+\mu)+\mu(\kappa+M^2)-\mu(q+\mu)^2]R_+R_-\} \nonumber\\ &+&\mu(\nu-2\kappa)[(\kappa+\mu^2)(R_+-R_-)(r_+-r_-) -\Sigma\sigma(R_++R_-)(r_++r_-)] \nonumber\\ &-&(\nu-2\Sigma\sigma-2\mu^2)[M\sigma(Q-\mu)-m\Sigma(q+\mu)](R_+r_--R_-r_+) \nonumber\\ &+&(\nu+2\Sigma\sigma-2\mu^2)[M\sigma(Q-\mu)+m\Sigma(q+\mu)](R_-r_--R_+r_+) \nonumber\\ &+&\Sigma[m\mu\nu+2M\mu\kappa+2R\kappa(q+\mu)] [2\sigma(r_++r_-)-R(r_+-r_-)] \nonumber\\ &+&\sigma[M\mu\nu+2m\mu\kappa-2R\kappa(Q-\mu)] [2\Sigma(R_++R_-)+R(R_+-R_-)] \nonumber\\ &+&\Sigma\sigma[\nu(Q-\mu)+2\kappa(q+\mu)] [2\sigma(r_+-r_-)-R(r_++r_-)] \nonumber\\ &+&\Sigma\sigma[\nu(q+\mu)+2\kappa(Q-\mu)] [2\Sigma(R_+-R_-)+R(R_++R_-)] \nonumber\\ &-&2\mu(\nu-2\kappa)\{\sigma[MR+\mu(Q-q-2\mu)](R_+-R_-) -\Sigma[mR+\mu(Q-q-2\mu)] \nonumber\\ &\times&(r_+-r_-)+\Sigma\sigma[(M-m)(r_++r_--R_+-R_-)+2(\nu+2\kappa)]\}. \label{A3_gen}  \eea

The magnetostatic analog of the double--Reissner--Nordstr\"om solution is now completely defined by formulas (\ref{EF_mag}), (\ref{EF})--(\ref{fg}) and (\ref{A3_gen}), $Q$ and $q$ being the individual magnetic charges of the constituents.

\section{Asymmetric black diholes}

The extreme and non--extreme black dihole solutions considered in the papers \cite{Emp} and \cite{ETe} are composed of two identical constituents and, as can be easily shown, they are symmetric with respect to the equatorial ($z=0$) plane. In this connection it would be of apparent interest to extend Emparan's notion of black dihole spacetimes to the equatorially non--symmetric case, and consider {\it the asymmetric black diholes} -- the systems of two black holes which possess unequal masses and charges equal in magnitude but opposite in sign. Let us substantiate the importance of asymmetric diholes by demonstrating below a remarkable general property shared by these configurations -- the possibility of balance of their black--hole constituents in the external field introduced via the Harrison transformation \cite{Har}. We shall treat this special balance effect within the framework of the magnetostatic case for which the application of the above--mentioned transformation is physically somewhat more transparent than in the case of electrostatics, but of course our conclusions will be equally valid for both cases in view of Bonnor's theorem \cite{Bon2}.

The action of the Harrison transformation on the magnetostatic analog of the double--Reissner--Nordstr\"om solution leads to the metric functions $\tilde f$, $\tilde\gamma$ and magnetic potential $\tilde A_\varphi$ of the form
\bea &&\tilde f=\lambda^2f, \quad e^{2\tilde\gamma}=\lambda^4 e^{2\gamma}, \quad \tilde A_\varphi=2B^{-1}[\lambda^{-1}(1+{\textstyle\frac{1}{2}}BA_\varphi)-1], \nonumber\\ &&\lambda=(1+{\textstyle\frac{1}{2}}BA_\varphi)^2+{\textstyle\frac{1}{4}}B^2\rho^2f^{-1}, \label{Har} \eea
where $f$, $\gamma$ and $A_\varphi$ are given by formulas (\ref{fg}), (\ref{A3_gen}) of the previous section, and $B$ is a real constant defining the exterior `uniform' magnetic field of the Bonnor--Melvin type \cite{Bon2,Mel}. When $B=0$, one recovers from (\ref{Har}) the magnetic version of the double--Reissner--Nordstr\"om solution; in this (asymptotically flat) case, equilibrium of two non--extreme charged black holes with positive masses does not seem possible \cite{PCo,BMA} because of an irremovable strut (conical deficit) on the part of the symmetry axis separating the constituents. The question is whether the strut can be removed for some $B\ne0$. We remind that the balance of aligned Reissner--Nordstr\"om black holes implies vanishing of the metric function $\gamma$ on all the parts of the symmetry axis outside the horizons, i.e., in our two--body case, the balance equations have the form \be \tilde\gamma^{(I)}=0, \quad \tilde\gamma^{(II)}=0, \quad \tilde\gamma^{(III)}=0, \label{bal_BM} \ee where the segments $I$, $II$ and $III$ of the symmetry axis are defined as $\rho=0, z>{\textstyle{\frac12}}R+\sigma$ (the upper part), $\rho=0, -{\textstyle{\frac12}}R+\Sigma<z<{\textstyle{\frac12}}R-\sigma$ (the intermediate part) and $\rho=0, z<-{\textstyle{\frac12}}R-\Sigma$ (the lower part), respectively. Taking into account (\ref{Har}), another way of writing (\ref{bal_BM}) is \be \lambda^4_{(I,II,III)}{e^{2\gamma}}^{(I,II,III)}=1, \label{bal_BM2} \ee where $\lambda_{(I,II,III)}$ and $\gamma^{(I,II,III)}$ stand for the values of $\lambda$ and $\gamma$ on the segments $I$, $II$, $III$ of the symmetry axis and refer to the `seed' double--Reissner--Nordstr\"om solution. From (\ref{fg}) and (\ref{EF}) follows that $\gamma^{(I,III)}=0$ identically, and also that $\rho^2f^{-1}|_{\rho=0}=0$; hence, the system of the balance equations finally takes the form \be A^{(I)}_\varphi=0, \quad (1+{\textstyle{\frac12}}BA^{(II)}_\varphi)^8 {e^{2\gamma}}^{(II)}=1, \quad A^{(III)}_\varphi=0, \label{bal_BM3} \ee $A^{(I,II,III)}_\varphi$ being the values of the magnetic potential (\ref{A3_gen}) on the respective parts of the $z$--axis. From (\ref{A3_gen}) we find the required axis values of the potential $A_\varphi$: \be A^{(I)}_\varphi\equiv 0, \quad A^{(II)}_\varphi=2q, \quad A^{(III)}_\varphi=2(Q+q), \label{A3_axis} \ee and it follows from (\ref{A3_axis}) that the first condition in (\ref{bal_BM3}) is satisfied automatically, while the fulfilment of the third condition in (\ref{bal_BM3}) is secured by vanishing of the total charge: \be Q_T\equiv Q+q=0. \label{cond3_sat} \ee Finding now $\gamma^{(II)}$ with the aid of (\ref{fg}) and (\ref{EF}), and solving the second equation in (\ref{bal_BM3}) for $B$, we arrive at the following value of the latter parameter at which two magnetically charged black holes will be balanced by the external magnetic field:
\be B=\frac{1}{q}\Biggl(\pm\sqrt[4]{\frac{\nu+2\kappa}{\nu-2\kappa}}-1 \Biggr), \quad Q+q=0. \label{B_value} \ee Note that for physical reasons ($B\to0$ when $R\to\infty$) one has to choose the upper sign in (\ref{B_value}). The zero total charge condition naturally singles out {\it the asymmetric black dihole} configurations as the ones whose black--hole constituents can achieve equilibrium in the external magnetic field, the precise `balance' value of such field being given by formulas (\ref{B_value}).

Once the significance of asymmetric black diholes is established, let us turn now to their more detailed description, returning again to the electrostatic picture and restricting our consideration to the asymptotically flat case (no external field). Then, the choice $q=-Q$ in the formulas (\ref{EF})--(\ref{nk}), (\ref{fg}) of the previous section permits us the introduction of the non--dimensional parameter $\mu$ in the expressions for $\Sigma$ and $\s$: \be \Sigma=\sqrt{M^2-Q^2(1-2\mu)}, \quad \s=\sqrt{m^2-Q^2(1-2\mu)}, \quad \mu:=\frac{M+m}{R+M+m}. \label{sig_as} \ee The expressions for $\nu$ and $\kappa$, with $q=-Q$, take the form \be \nu=R^2-M^2-m^2+2Q^2(1-\mu)^2, \quad \kappa=Mm+Q^2(1-\mu)^2, \label{nk_as} \ee while for the electric potential $A_t$ and metric functions $f$ and $\gamma$ we obtain \bea
f&=&\frac{\A^2-4\B^2+4Q^2\C^2}{(\A+2\B)^2}, \quad A_t=-\frac{2Q\C}{\A+2\B}, \nonumber\\
e^{2\gamma}&=&\frac{\A^2-4\B^2+4Q^2\C^2}{K_0^2 R_+R_-r_+r_-}, \quad K_0=4\Sigma\,\s[R^2-(M-m)^2+4Q^2(1-\mu)^2], \nonumber\\ \A&=&\Sigma\,\s[\nu(R_++R_-)(r_++r_-)+4\kappa(R_+R_-+r_+r_-)] \nonumber\\ &-&(Q^2\mu^2\nu-2\kappa^2)(R_+-R_-)(r_+-r_-), \nonumber\\ \B&=&\Sigma\,\s[(m\nu +2M\kappa)(R_++R_-)+(M\nu+2m\kappa)(r_++r_-)] \nonumber\\ &+&Q^2(\mu-\mu^2)(\nu-2\kappa)[\s(R_+-R_-)-\Sigma(r_+-r_-)] \nonumber\\ &-&2R\kappa[M\s(R_+-R_-)-m\Sigma(r_+-r_-)], \nonumber\\  \C&=&\Sigma\,\s(1-\mu)(\nu-2\kappa)(r_++r_--R_+-R_-) \nonumber\\ &+&\s[M\mu\nu +2\kappa(m\mu-R+R\mu)](R_+-R_-) \nonumber\\ &+&\Sigma[m\mu\nu+2\kappa(M\mu-R+R\mu)](r_+-r_-), \nonumber\\ R_\pm&=&\sqrt{\rho^2+(z+{\textstyle\frac{1}{2}}R\pm\Sigma)^2}, \quad r_\pm=\sqrt{\rho^2+(z-{\textstyle\frac{1}{2}}R\pm\s)^2}. \label{fg_as} \eea

From the above formulas (\ref{sig_as}) and (\ref{nk_as}) one immediately draws the following three interesting conclusions about the properties of asymmetric black diholes:

($i$) Since $\Sigma$ and $\s$ cannot take the same values for different positive values of $M$ and $m$, the asymmetric diholes with two extreme components ($\Sigma=\s=0$) do not exist.

($ii$) The formula for the interaction force between the black--hole constituents in an asymmetric dihole is readily obtainable from the general expression derived in \cite{Man} for the double--Reissner--Nordstr\"om solution and from (\ref{nk_as}); it has the form
\be {\cal F}=\frac{\kappa}{\nu-2\kappa}=\frac{Mm+Q^2(1-\mu)^2}{R^2-(M+m)^2}, \label{force} \ee which means that no equilibrium states without a supporting strut between the constituents (such states are defined by the equation ${\cal F}=0$) can be achieved, at finite separation, for arbitrary values of $Q$ and any positive values of $M$ and $m$.

($iii$) The absolute value of charge $Q$ in the asymmetric black diholes can considerably exceed the values of $M$ and $m$ due to the presence of the factor $(1-2\mu)$ in the expressions of $\Sigma$ and $\s$.

The physical quantities $\Phi_\Sigma$, $\Phi_\sigma$, $A_\Sigma$ and $A_\sigma$ characterizing the asymmetric diholes have a somewhat simpler form than in the case of the general double--Reissner--Nordstr\"om solution. Thus, for $\Phi_\Sigma$ and $\Phi_\sigma$ we have \be \Phi_\Sigma=\frac{Q(1-2\mu)}{M+\Sigma}, \quad \Phi_\sigma=-\frac{Q(1-2\mu)}{m+\sigma}, \label{fi} \ee and for $A_\Sigma$ and $A_\sigma$ we get from (\ref{ai_gen}) \be A_\Sigma=\frac{4\pi(R+M+m)^2(M+\Sigma)^2}{(R+\Sigma)^2-\sigma^2}, \quad A_\sigma=\frac{4\pi(R+M+m)^2(m+\sigma)^2}{(R+\sigma)^2-\Sigma^2}, \label{ai} \ee so that the charge $Q$ is not present explicitly in (\ref{ai}). Mention that the electric dipole moment of the asymmetric dihole is equal to $-Q(R-M-m)$.

In the limit of equal constituents ($\Sigma=\sigma$, $M=m$), one easily recovers from (\ref{fi}) and (\ref{ai}) the expressions for the electric potential and horizon's area found by Emparan and Teo for a non--extreme symmetric dihole \cite{ETe}.

In conclusion of this section we write out explicitly the expression of the potential $A_\varphi$ for the magnetic analog of the electrically charged asymmetric black dihole [apparently, the functions $\A$, $\B$ and $\C$ of this analog are the same as in formulas (\ref{fg_as})]:
\bea A_\varphi&=&\frac{Q(\II-2z\C)}{\A+2\B}, \nonumber\\ \II&=&4\Sigma\sigma[R(1-\mu)-\mu(M+m)](MR_+R_-+mr_+r_-) \nonumber\\ &+&\mu(\nu-2\kappa)[(\kappa+\mu^2Q^2)(R_+-R_-)(r_+-r_-) -\Sigma\sigma(R_++R_-)(r_++r_-)] \nonumber\\ &-&(1-\mu)\{[R^2-(\Sigma+\sigma)^2](M\sigma+m\Sigma)(R_+r_--R_-r_+) \nonumber\\ &+&[R^2-(\Sigma-\sigma)^2](M\sigma-m\Sigma)(R_+r_+-R_-r_-)\} \nonumber\\ &+&\Sigma[m\mu\nu+2M\mu\kappa-2R\kappa(1-\mu)] [2\sigma(r_++r_-)-R(r_+-r_-)] \nonumber\\ &+&\sigma[M\mu\nu+2m\mu\kappa-2R\kappa(1-\mu)] [2\Sigma(R_++R_-)+R(R_+-R_-)] \nonumber\\ &-&\Sigma\sigma R(1-\mu)(\nu-2\kappa) (R_++R_-+r_++r_-) \nonumber\\ &-&2(\nu-2\kappa)\{\sigma[MR\mu+(\Sigma^2+2Q^2\mu^2)(1-\mu)](R_+-R_-) \nonumber\\ &-&\Sigma[mR\mu+(\sigma^2+2Q^2\mu^2)(1-\mu)](r_+-r_-) \nonumber\\ &+&\mu\Sigma\sigma[(M-m)(r_++r_--R_+-R_-)+2(\nu+2\kappa)]\}, \label{A3_dihole}  \eea
and we also remark that the explicit form of formula (\ref{B_value}) for the parameter $B$ defining equilibrium of the magnetically charged black--hole constituents in the external magnetic field, with account of (\ref{sig_as}) and (\ref{nk_as}), takes the form
\be B=-\frac{1}{Q}\Biggl(\sqrt[4]{\frac{R^2-(M-m)^2+4Q^2(1-\mu)^2} {R^2-(M+m)^2}}-1 \Biggr), \ee $Q$ being the magnetic charge of the lower constituent.

\section{The double--Gibbons--Maeda solution}

To generalize the double--Reissner--Nordstr\"om solution to the Einstein--Maxwell--dilaton theory arising from the Lagrangian \cite{GMa,GHS} \be {\cal L}={1\over 16\pi}\sqrt{-g}\,\,[R-2(\nabla\phi)^2-e^{-2\a\phi}F^2], \label{lagr} \ee where $\phi$ is the dilaton field and $\a$ the arbitrary coupling constant ($\a=0$ for the pure
Einstein--Maxwell fields, $\a=1$ for the low
energy effective limit of string theory and $\a=\sqrt{3}$ for the Kaluza--Klein theory), it is advantageous to make use of the solution--generating procedure described in the papers \cite{Emp,ETe}. Recall that according to this procedure, given a seed electrostatic solution $f$, $\gamma$ and $A_t$, its dilatonic generalization $\widehat f$, $\widehat\gamma$, $\widehat{A}_t$ and $\phi$ is defined by the formulas \bea \widehat f&=&f^{\frac{1}{1+\a^2}} e^{-2\a\phi_0}, \quad \widehat\gamma=\frac{1}{1+\a^2}\gamma+\gamma_0, \nonumber\\ \widehat A_t&=&\frac{1}{\sqrt{1+\a^2}}A_t, \quad e^{2\phi}=f^{\frac{\a}{1+\a^2}} e^{2\phi_0}, \label{GF} \eea where $\widehat f$ and $\widehat\gamma$ enter the line element \be
\d s^2=\widehat f^{-1}[e^{2\widehat\gamma}(\d\rho^2+\d z^2)+\rho^2\d\varphi^2]-\widehat f\,\d
t^2, \label{Weyl2} \ee $\phi_0$ is an arbitrary solution of the equation \be \phi_{0,\rho,\rho}+\rho^{-1}\phi_{0,\rho}+\phi_{0,z,z}=0, \label{L_eq} \ee and $\gamma_0$ is obtainable from $\phi_0$ by solving the system \bea \gamma_{0,\rho}&=&(1+\a^2)\rho(\phi_{0,\rho}^2-\phi_{0,z}^2), \nonumber\\
\gamma_{0,z}&=&2(1+\a^2)\rho\phi_{0,\rho}\phi_{0,z}, \label{g_sys} \eea the integrability condition of which is equation (\ref{L_eq}).

In the case of the double--Reissner--Nordstr\"om solution the functions $f$ and $\gamma$ are given by formulas (\ref{fg}), while the potential $A_t$ is equal to $-\Phi$ from (\ref{EF}). Therefore, there only remains to identify the function $\phi_0$ and the corresponding $\gamma_0$ to obtain the desired generalization. Since our $\phi_0$ has in particular to generalize the Emparan and Teo's analogous choice for the symmetric black dihole solution \cite{ETe}, we have to take as $\phi_0$ the linear superposition of two arbitrary Schwarzschild solutions with the ``masses'' $\Sigma$ and $\s$ and find the corresponding $\gamma_0$ using the results of the paper \cite{IKh}. Then, after the application of the procedure (\ref{GF}) to the double--Reissner--Nordstr\"om solution we finally obtain \bea \widehat f&=&\biggl[\frac{\A^2-\B^2+\C^2} {(\A+\B)^2}\biggr]^{\frac{1}{1+\a^2}}e^{-2\a\phi_0}, \nonumber\\ e^{2\widehat\gamma}&=&\biggl[\frac{\A^2-\B^2+\C^2} {K_0^2R_+R_-r_+r_-}\biggr]^{\frac{1}{1+\a^2}}e^{2\gamma_0},  \nonumber\\ \widehat{A}_t&=&-\frac{1}{\sqrt{1+\a^2}}\, \frac{\C}{\A+\B}, \nonumber\\ e^{2\phi}&=&\biggl[\frac{\A^2-\B^2+\C^2} {(\A+\B)^2}\biggr]^{\frac{\a}{1+\a^2}}e^{2\phi_0}, \label{MFD} \eea where \bea e^{-2\phi_0}&=&\biggl(\frac{R_++R_--2\Sigma}{R_++R_-+2\Sigma}\cdot \frac{r_++r_--2\s}{r_++r_-+2\s}\biggr)^{\frac{\a}{1+\a^2}}, \nonumber\\ e^{2\gamma_0}&=& \biggl[\frac{R_+R_-+\rho^2+(z+{\textstyle\frac12}R)^2-\Sigma^2}{2R_+R_-} \nonumber\\ &\times&\frac{r_+r_-+\rho^2+(z-{\textstyle\frac12}R)^2-\s^2}{2r_+r_-} \nonumber\\ &\times&\frac{R_+r_-+\rho^2+(z+{\textstyle\frac12}R+\Sigma)(z-{\textstyle\frac12}R-\s)} {R_-r_-+\rho^2+(z+{\textstyle\frac12}R-\Sigma)(z-{\textstyle\frac12}R-\s)} \nonumber\\ &\times&\frac{R_-r_++\rho^2+(z+{\textstyle\frac12}R-\Sigma)(z-{\textstyle\frac12}R+\s)} {R_+r_++\rho^2+(z+{\textstyle\frac12}R+\Sigma)(z-{\textstyle\frac12}R+\s)} \biggr]^{\frac{\a^2}{1+\a^2}}. \label{f0} \eea

This completes the construction of the dilatonic generalization of the electrostatic metric for two charged black holes. In the limit of identical black--hole constituents with opposite electric charges ($\Sigma=\s$, $q=-Q$), from the formulas obtained one easily recovers the Emparan--Teo non--extremal dihole solution \cite{ETe} in the physical parametrization of paper \cite{CCM}. Moreover, by setting either $m=q=0$ or $M=Q=0$ in the above formulas we arrive at the well--known Gibbons--Maeda solution \cite{GMa} for a single Reissner--Nordstr\"om black hole in the 4--dimensional dilaton gravity. So, observing a certain parallelism with the well--known double--Kerr solution \cite{KNe}, it would be most appropriate to call the dilaton two--body configuration considered in this section {\it the double--Gibbons--Maeda solution}.

The presence of the dilaton field in the double--Gibbons--Maeda solution only slightly changes the form of the corresponding potentials $\Phi_\Sigma$ and $\Phi_\sigma$ in which the additional factor $(1+\a^2)^{-1/2}$ arises compared to the expressions (\ref{fi_gen}) of the double--Reissner--Nordstr\"om solution. On the other hand, the dilaton field effects more substantially the form of the horizons' areas, $A_\Sigma$ and $A_\sigma$, due to the presence of the functions $\phi_0$ and $\gamma_0$ in the expressions for $\widehat f$ and $\widehat\gamma$. Nonetheless, for $A_\Sigma$ and $A_\sigma$ we have been able to obtain the following compact expressions, namely, \bea A_\Sigma&=&\frac{4\pi(4\Sigma^2)^{\frac{\a^2}{1+\a^2}} [(R+M+m)(M+\Sigma)-Q(Q+q)]^{\frac{2}{1+\a^2}}} {(R+\Sigma-\sigma)(R+\Sigma+\sigma)^{\frac{1-\a^2}{1+\a^2}}}, \nonumber\\ A_\sigma&=&\frac{4\pi(4\sigma^2)^{\frac{\a^2}{1+\a^2}} [(R+M+m)(m+\sigma)-q(Q+q)]^{\frac{2}{1+\a^2}}} {(R-\Sigma+\sigma)(R+\Sigma+\sigma)^{\frac{1-\a^2}{1+\a^2}}}, \label{ai_gen_dil} \eea which generalize formulas (\ref{ai_gen}) in a nice manner. Notice that, as expected, in the presence of the non--zero dilaton field ($\a\ne0$) the extremal dilatonic holes ($\Sigma=0$ or/and $\sigma=0$) develop null singularities.

For the interaction energy in the double--Gibbons--Maeda solution we find the expression \be V_{\rm int}=\frac{(R-M-m)\{ (\nu-2\kappa)^{\frac{1-\a^2}{1+\a^2}} [R^2-(\Sigma+\sigma)^2]^{\frac{2\a^2}{1+\a^2}}-(\nu+2\kappa)\}} {4(\nu+2\kappa)}, \label{energy} \ee which follows from the formula \cite{ETe} \be V_{\rm int}=\frac{L}{4}(e^{\widehat\gamma_0}-1), \ee $L$ being the coordinate length of the strut, and $\widehat\gamma_0$ being the value of the metric function $\widehat\gamma$ on the part of the symmetry axis separating the black--hole constituents.

We end up this section by observing that the magnetic analog of the double--Gibbons--Maeda solution is defined by a slight modification of the generating formulas (\ref{GF}), in which one only has to replace the electric potential $A_t$ by the magnetic potential $A_\varphi$ defined by (\ref{A3_gen}) and set $\phi$ to $-\phi$; the choice of the functions $f$, $\gamma$, $\phi_0$ and $\gamma_0$ is the same as in the electrostatic solution.

\section{Conclusions}

In the present paper we have considered several applications of the double--Reissner--Nordstr\"om solution to the binary systems of charged black holes. We started from a brief review of the electrostatic case, supplementing it with a derivation of a remarkably simple formulas for horizons' areas and surface gravities of two interacting Reissner--Nordstr\"om black holes. We then obtained a magnetic analog of the double--Reissner--Nordstr\"om solution which required the construction of the corresponding magnetic potential $A_\varphi$ with the aid of Sibgatullin's integral method. The analysis of the behavior of magnetically charged black holes in the external magnetic field naturally led us to {\it asymmetric black diholes} as an important family of binary configurations for which equilibrium of the black--hole constituents is achievable, and we found a precise value of the parameter $B$ of the external field at which the balance of the gravitational and magnetic forces occurs. One would anticipate that the `balanced' asymmetric black dihole spacetimes could find interesting applications in the numerical simulations of the non--stationary black--hole configurations as the initial static conditions. The absence of asymmetric dihole configurations with two extreme components is a curious fact established in this paper which lends additional importance to the Bonnor magnetic dipole metric \cite{Bon} and to some of its generalizations \cite{Emp} as the extreme black dihole models. At the same time, it should be mentioned that our paper provides one with a basis for the analysis of the black dihole configurations involving one extreme and one non--extreme components.

There are various reasons for thinking that the double--Gibbons--Maeda solution, obtained and briefly discussed in section~IV, opens new promising opportunities for the study of binary black--hole configurations in the presence of a dilaton field. First of all, this solution describes a pair of {\it arbitrary} Gibbons--Maeda black holes and, therefore, permits one to study more general binary black--hole systems, with zero or non--zero total charge, than earlier considered in the literature; and it is important that the existence of simple formulas (\ref{ai_gen_dil}) for the horizon areas strongly suggests that the overall physical analysis of the double--Gibbons--Maeda spacetime is feasible as well. Furthermore, the availability of the general expression for the magnetic potential $A_\varphi$ now makes it possible a straightforward application of the dilatonic Harrison transformation \cite{DGK} to the solution obtained and searching for equilibrium configurations of the constituents in the framework of dilatonic asymmetric black diholes surrounded by the external magnetic field, in analogy with the pure Einstein--Maxwell case. In this respect it is worth mentioning that in the paper \cite{ETe} a microscopic description of the entropy of interacting equal black holes carrying opposite charges was given using the `effective string' model and the large separation approximation needed for obtaining a weak strut singularity. Although it is clear that a similar analysis is also applicable to our asymmetric black diholes, it seems more advantageous to develop a description which would be not subject to the restriction of large separation; for this purpose, as was already observed in \cite{ETe}, the well--behaved black diholes balanced by the external field are the best solutions to employ. Since the crucial point in the latter solutions is the explicit form of the magnetic potential $A_\varphi$ which was constructed in the present paper, we expect that the desired improved microscopic description of the entropy will be provided in the future.

\section*{Acknowledgments}

We would like to thank Eloy Ay\'on--Beato for interesting and helpful discussions. We are also thankful to Erasmo G\'omez for technical computer assistance. This work was partially supported by Project 45946--F from CONACyT, Mexico, by Project FIS2006--05319 from MCyT, Spain, and by Project SA010C05 from Junta de Castilla y Le\'on, Spain.

\newpage

\begin{figure}[htb]
\centerline{\epsfysize=90mm\epsffile{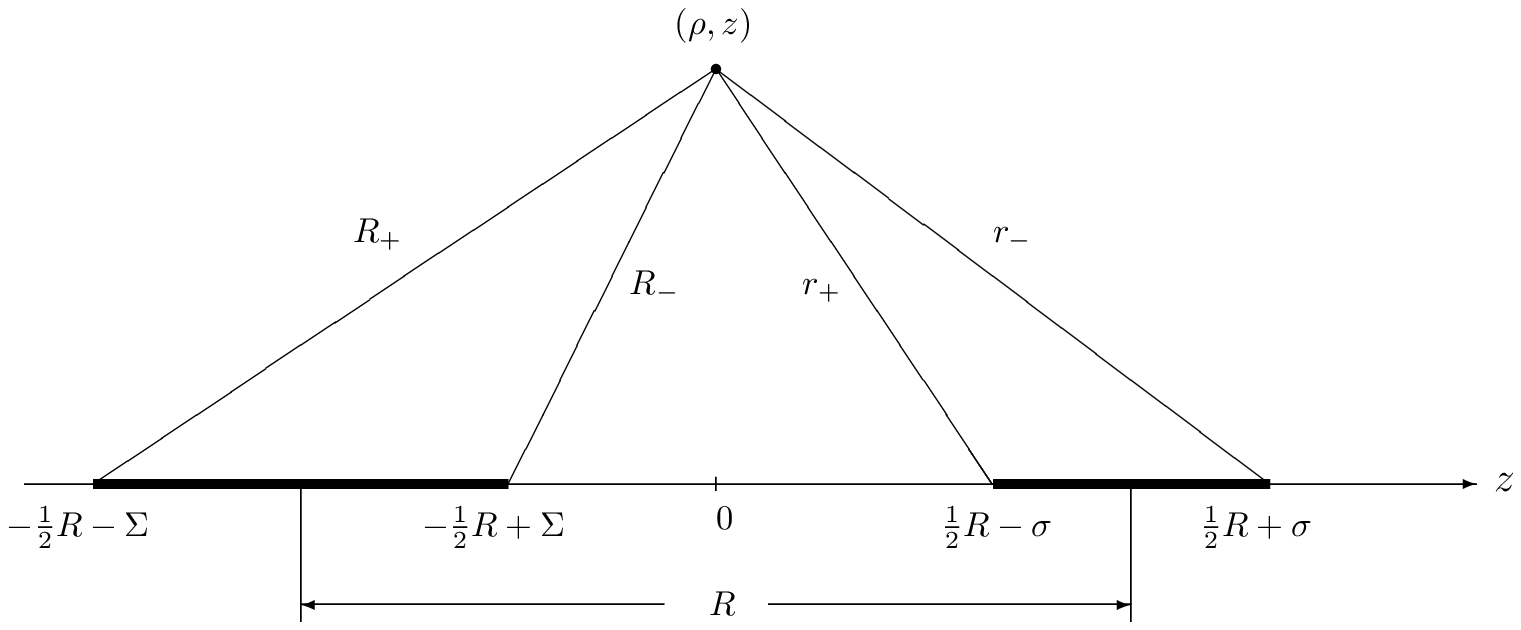}} \caption{Location of the black--hole constituents on the symmetry axis.}
\end{figure}


\begin{references}
\bibitem{GMa} G. W. Gibbons and K. Maeda, Nucl. Phys. B
{\bf 298}, 741 (1988).
\bibitem{DGe} A.~Davidson and E.~Gedalin, \J{\PLB}{339}{304}{1994}.
\bibitem{Emp} R.~Emparan, \J{\PRD}{61}{104009}{2000}.
\bibitem{LTe} Y.~C.~Liang and E.~Teo, \J{\PRD}{64}{024019}{2001}.
\bibitem{ETe} R.~Emparan and E.~Teo, \J{\NPB}{610}{190}{2001}.
\bibitem{Bon} W.~B.~Bonnor, \J{\ZP}{190}{444}{1966}.
\bibitem{CCM} J. A.~C\'azarez, H. Garc\'\i a--Compe\'an and V.~S.~Manko, \J{\PLB}{662}{213}{2008}; \J{\ib}{665}{426}{2008}.
\bibitem{EMR} F.~J.~Ernst, V.~S.~Manko and E.~Ruiz, \J{\CQG}{23}{4945}{2006}.
\bibitem{Rei} H.~Reissner, \J{\APL}{50}{106}{1916}.
\bibitem{Nor} G.~Nordstr\"om, Proc. K. Ned. Akad. Wet. {\bf 20}, 1238 (1918).
\bibitem{BMA} N.~Bret\'on, V.~S.~Manko and J.~Aguilar--S\'anchez, \J{\CQG}{15}{3071}{1998}.
\bibitem{Man} V.~S.~Manko, \J{\PRD}{76}{124032}{2007}.
\bibitem{Bon2} W.~B.~Bonnor, Proc. Phys. Soc. Lond. A {\bf 67}, 225 (1954).
\bibitem{Mel} M.~A.~Melvin, Phys. Lett. {\bf 8}, 65 (1964).
\bibitem{CMRS} B. Chng, R. Mann, E. Radu and C. Stelea, \J{\JHEP}{12}{009}{2008}.
\bibitem{Har} B. K. Harrison, \J{\JMP}{9}{1744}{1968}.
\bibitem{VCh} G. G.~Varzugin and A. S.~Chistyakov, Class. Quantun Grav. {\bf 19}, 4553 (2002).
\bibitem{Kom} A.~Komar, Phys. Rev. {\bf 113}, 934 (1959).
\bibitem{Ern} F.~J.~Ernst, Phys. Rev. {\bf 168}, 1415 (1968).
\bibitem{Sma} L.~Smarr, \J{\PRL}{30}{71}{1973}.
\bibitem{Kin} W. Kinnersley, \J{\JMP}{18}{1529}{1977}.
\bibitem{Sib} N.~R.~Sibgatullin, {\it Oscillations and Waves in
Strong Gra\-vitational and Electromagnetic Fields} (Nauka, Moscow,
1984) [English translation (Springer--Verlag, Berlin, 1991)]; V.~S.~Manko and N.~R.~Sibgatullin, \J{\CQG}{10}{1383}{1993}.
\bibitem{RMM} E.~Ruiz, V.~S.~Manko and J.~Mart\'\i n,
\J{\PRD}{51}{4192}{1995}.
\bibitem{PCo} G.~P.~Perry and F.~I.~Cooperstock,
\J{\CQG}{14}{1329}{1997}.
\bibitem{GHS} D. Garfinkle, G. T. Horowitz and A. Strominger,
Phys. Rev. D {\bf 43}, 3140 (1991); {\it ibid.} {\bf 45}, 3888 (1992), Erratum.
\bibitem{IKh} W.~Israel and K. A. Khan, \J{\NC}{33}{331}{1964}.
\bibitem{KNe} D. Kramer and G. Neugebauer, \J{\PLA}{75}{259}{1980}.
\bibitem{DGK} F. Dowker, J. P. Gauntlett, D. A. Kastor and J. Traschen, \J{\PRD}{49}{2009}{1999}.

\end{references}
\end{document}